\documentclass[preprint,12pt]{elsarticle}

\usepackage{amssymb}
\usepackage{amsmath}

\usepackage{lineno}

\usepackage{diagbox}
\graphicspath{{figures/}}

\journal{Computational Materials Sciences}

\begin{document}
		\begin{frontmatter}
			
			
			
			\title{Predicting parameters of a model cuprate superconductor using machine learning} 
			
			\author{V.~A.~Ulitko}
			\ead{vasiliy.ulitko@urfu.ru}
			\author{D.~N.~Yasinskaya}
			\author{S.~A.~Bezzubin}
			\author{A.~A.~Koshelev}
			\author{Y.~D.~Panov}
			\affiliation{organization={Ural Federal University},
				addressline={19 Mira street},
				city={Ekaterinburg},
				postcode={620002},
				country={Russia}}
			
			\begin{abstract}
				The computational complexity of calculating phase diagrams for multi-parameter models significantly limits the ability to select parameters that correspond to experimental data.  
				This work presents a machine learning method for solving the inverse problem -- forecasting the parameters of a model Hamiltonian for a cuprate superconductor based on its phase diagram.
				A comparative study of three deep learning architectures was conducted: VGG, ResNet, and U-Net.
				The latter was adapted for regression tasks and demonstrated the best performance.
				Training the U-Net model was performed on an extensive dataset of phase diagrams calculated within the mean-field approximation, followed by validation on data obtained using a semi-classical heat bath algorithm for Monte Carlo simulations.
				It is shown that the model accurately predicts all considered Hamiltonian parameters, and areas of low prediction accuracy correspond to regions of parametric insensitivity in the phase diagrams. 
				This allows for the extraction of physically interpretable patterns and validation of the significance of parameters for the system.  
				The results confirm the promising potential of applying machine learning to analyze complex physical models in condensed matter physics.
			\end{abstract}
			
			
			
			
		\end{frontmatter}
		
			
			
			\section*{Introduction}
			\label{sec1}
			
			Modern microscopic models of complex materials with non-trivial properties are typically multi-parametric, and the direct calculation of their phase diagrams is computationally expensive.
			The problem of computational complexity becomes especially critical when solving the inverse problem, which involves finding the model parameters that best reproduce an experimentally observed effect or a phase diagram.
			
			An example of such a multi-parametric system is the previously proposed~\cite{Moskvin2011PRB, Moskvin2013JPCP, Moskvin2022JMMM} pseudospin model of high-temperature superconducting (HTSC) cuprate.
			For this model, we derived equations for the critical temperatures of various ordered phases in mean-field approximation (MFA)~\cite{Panov2019PMM} and constructed temperature phase diagrams~\cite{Moskvin2020PSS, Moskvin2021CM} -- dependencies of the phase state on temperature and consentation.
			The possibility of simulating this system using the classical Monte Carlo method was also discussed in~\cite{Panov2021MC,Ulitko2023classical}.
			
			Due to the large number of independent model parameters, solving the inverse problem of finding a parameter set that reproduces the experimental phase diagrams for such a system is extremely challenging.
			In this regard, it is feasible to explore the possibility of applying machine learning to reduce the required number of phase diagram calculations.
			
			Previously, machine learning methods have been used to predict phase boundaries on experimentally and theoretically determined phase diagrams~\cite{lund2022machine,casert2019interpretable}, to identify stable phases and predict phase diagrams~\cite{zipoli2022prediction,bayo2025machine,aghaaminiha2020machine,sun2024precise}, and to study phase transitions~\cite{van2017learning,zhang2019machine,canabarro2019unveiling}.
			Specialized neural networks have been developed for analyzing images of magnetic domains to quantitatively estimate Hamiltonian parameters~\cite{kwon2020magnetic}, while convolutional networks have successfully constructed detailed phase diagrams for systems with skyrmions and other topological phases~\cite{gomez2022machine,che2020topological}.
			In the context of HTSC research, machine learning has been successfully applied to predict critical temperatures of various systems~\cite{stanev2018machine,zhang2022predicting,xie2022machine,nieto2024predicting}.
			In this work, we developed and tested a machine learning method based on a U-Net convolutional neural network to solve the inverse problem: predicting the parameters of a model HTSC cuprate from its theoretical phase diagram.
			The primary goal of this work is to overcome the computational complexity of calculating phase diagrams and to develop an efficient method for predicting model parameters.
			This research focuses not only on demonstrating the principal feasibility of applying machine learning but also on analyzing the physical interpretability of the obtained results.
			
			The article is organized as follows.
			The first section introduces the physical model of the HTSC cuprate and the modeling methodology.
			The second section  compares different learning architectures and discusses their features.
			The third section presents the results of numerical experiments conducted using the U-Net architecture to predict model parameters.
			
			\section*{Physical model and modeling methods}
			
			The rich phase diagram of HTSC cuprates makes the direct statistical modeling of these systems within an effective pseudospin framework highly relevant~\cite{Moskvin2022JMMM}.
			
			The fundamental structural elements of HTSC cuprates are the CuO$_2$ planes, composed of square CuO$_4$ clusters connected via shared oxygen atoms.
			Within the effective pseudospin model for cuprates, the basis of the local Hilbert space for a CuO$_4$ cluster is chosen as a quartet of states $\left| 1 M s m \right\rangle$.
			In the effective pseudospin model for cuprates, the basis of the local Hilbert space formed by the quartet of states $\left| 1 M s m \right\rangle$. 
			The projections $M = {-}1,\, 0,\, {+}1$ of pseudospin $S=1$ correspond to three multi-electronic states of the cluster [CuO$_4$]$^{7-,6-,5-}$.
			States with $M = \pm1$ are spin singlets ($s=0$), while the state with $M=0$ is a spin doublet with $s=1/2$.
			
			The Hamiltonian of the model cuprate can be written as follows:  
			\begin{equation}
				\hat{H} 
				= \hat{H}_{ch} 
				+ \hat{H}_{exc} 
				+ \hat{H}_{tr}^{(1)} 
				+ \hat{H}_{tr}^{(2)} 
				- \mu \sum_i \hat{S}_{z,i} .
				\label{eq:Ham0}
			\end{equation}
			The first term
			\begin{equation}
				\hat{H}_{ch} =
				\Delta \sum_i \hat{S}_{z,i}^2 
				+ V \sum_{\left\langle ij \right\rangle} \hat{S}_{z,i} \hat{S}_{z,j} 
			\end{equation}
			describes local (with $\Delta=U/2$, where $U$ is the correlation parameter) and non-local ($V>0$) density-density correlations. 
			Here, the operator $\hat{S}_{z}$ represents the z-component of the pseudospin $S=1$. 
			The summation runs over $N$ sites of a square lattice, and $\left\langle ij \right\rangle$ denotes nearest neighbors.
			The operator 
			\begin{equation}
				\hat{H}_{exc}
				=
				Js^2 \sum_{\langle ij \rangle} \hat{\boldsymbol{\sigma}}_i \hat{\boldsymbol{\sigma}}_j 
			\end{equation}
			describes the antiferromagnetic ($J>0$) Heisenberg exchange interaction for [CuO]$_4^{6-}$ centers, 
			where operators $\hat{\boldsymbol{\sigma}}=\hat{P}_0 \, \hat{\mathbf{s}}/s$ account for the spin density on the site, with $\hat{P}_0 = 1-\hat{S}_z^2$, 
			and $\hat{\mathbf{s}}$ is the spin operator with $s=1/2$ \cite{Panov2016JSNM}. 
			The operator
			\begin{multline}
				\hat{H}_{tr}^{(1)}
				\;=\;
				- \sum_{\left\langle ij\right\rangle m} 
				\Big[  
				t_p \hat{P}_{m,i}^{{+}} \hat{P}_{m,j}^{}  +  t_n \hat{N}_{m,i}^{{+}} \hat{N}_{m,j}^{} 
				\\
				{} + 
				\frac{t_{pn}}{2} \big( \hat{P}_{m,i}^{{+}} \hat{N}_{m,j}^{} + \hat{N}_{m,i}^{{+}} \hat{P}_{m,j}^{} \big) 
				\Big]
				+ h.c. 
			\end{multline}
			describes three types of correlated single-particle transport~\cite{Moskvin2011PRB,Moskvin2013JPCP}:
			terms proportional to $t_p$ ($t_n$) describe the transport of [CuO]$_4^{5-}$ ([CuO]$_4^{7-}$) centers across the [CuO]$_4^{6-}$ lattice;  
			terms proportional to $t_{pn}$ correspond to disproportionation and recombination processes: 
			$[\mathrm{CuO}]_4^{6-}+[\mathrm{CuO}]_4^{6-} \longleftrightarrow [\mathrm{CuO}]_4^{7-}+[\mathrm{CuO}]_4^{5-}$. 
			The spin projection $m=\pm1/2$ refers to the $s=1/2$ spin states;,  
			the orbital parts of the operators $\hat{P}$ and $\hat{N}$ are expressed via the $S=1$ pseudospin operators:
			$\hat{P}^{+} = \big(\hat{S}_{+} + \hat{T}_{+}\big)/2$, 
			$\hat{N}^{+} = \big(\hat{S}_{+} - \hat{T}_{+}\big)/2$, 
			$\hat{T}_{+} = \hat{S}_{z} \hat{S}_{+} + \hat{S}_{+} \hat{S}_{z}$, 
			$\hat{S}_{+} = \big( \hat{S}_{x} + i \hat{S}_{y} \big) / \sqrt{2}$.
			The operator
			\begin{equation}
				\hat{H}_{tr}^{(2)}
				=	- t_b \sum_{\left\langle ij\right\rangle} 
				\left( \hat{S}_{{+}i}^2 \hat{S}_{{-}j}^2 + \hat{S}_{{+}j}^2 \hat{S}_{{-}i}^2 \right)
			\end{equation}
			describes two-particle transport (transport of a local composite boson~\cite{Moskvin2011PRB,Moskvin2013JPCP}): 
			$[\mathrm{CuO}]_4^{5-}+[\mathrm{CuO}]_4^{7-} \leftrightarrow [\mathrm{CuO}]_4^{7-}+[\mathrm{CuO}]_4^{5-}$.
			The last sum with the chemical potential $\mu$ allows us to take into account the condition of constant charge density:
			
			\begin{equation}
				n = \left\langle \sum_i \hat{S}_{z,i} \right\rangle / N = const.
			\end{equation}
			
			In the mean-field approximation (MFA), equations for the critical temperatures of ordered phases were derived in \cite{Panov2019PMM}, considering cases where only one of the possible order parameters is non-zero.
			The charge-ordered (CO) phase is characterized by a non-zero order parameter $L = \big( S_{z,A} - S_{z,B} \big) / 2$, where $S_{z,A}$ and $S_{z,B}$ are the statistical averages of the $\hat{S}{z}$ operator on sublattices $A$ and $B$, which describe a checkerboard subdivision of the square lattice constructed from $[\mathrm{CuO}]_4$ clusters.
			Analogous to the model of local (hard-core) bosons~\cite{Micnas1990RMP}, the phase with a non-zero average $\big\langle \hat{S}_{+}^2 \big\rangle$ can be termed a Bose superfluid (BS) phase.
			For the antiferromagnetic (AFM) phase, the order parameter is the magnitude of the antiferromagnetic vector $\mathbf{l} = \big( \boldsymbol{\sigma}_{A} - \boldsymbol{\sigma}_{B} \big) / 2$, where $\boldsymbol{\sigma}_{A}$ and $\boldsymbol{\sigma}_{B}$ are the statistical averages of the spin operators ($\hat{\boldsymbol{\sigma}}$) on sublattices $A$ and $B$.
			In phases with non-zero order parameters $\big\langle \hat{P}_{m}^{+} \big\rangle$ (P phase) and $\big\langle \hat{N}_{m}^{+} \big\rangle$ (N phase), correlated single-particle transfer of hole (P) or electron (N) type is realized.
			
			Numerous experimental data indicate the key role of the competition between different phases and the formation of complex phase states in HTSC cuprates~\cite{Fradkin2015}.
			Investigating these issues with analytical methods remains challenging; however, within the effective pseudospin model, it is possible to address this problem using numerical statistical modeling techniques.
			
			For numerical modeling, we use the heat bath Monte Carlo algorithm~\cite{Miyatake1986}, which we previously applied to model orthonickelates~\cite{Panov2024FTT}.
			A brief description of the algorithm for the HTSC cuprate model \eqref{eq:Ham0} is provided below; details can be found in the preprint~\cite{Panov2025modified}.
			The system's wave function is written as a product of the wave functions on individual clusters:
			
			\begin{equation}
				\left| \Psi \right\rangle = \prod_{\nu} \left| \psi_{\nu} \right\rangle , \;\;
				\left| \psi_{\nu} \right\rangle 
				= \sum_{Mm} a_{Mm}^{\nu} \left| 1Msm \right\rangle .
				\label{eq:Psi}
			\end{equation}
			
			This allows for the construction of the Hamiltonian for a single cluster $\nu$ by averaging over the states of all other clusters $\mu \neq \nu$:
			
			\begin{equation}
				\hat{H}_{\nu} = \big\langle \tilde{ \Psi }_{\nu} \big| \hat{H} \big| \tilde{ \Psi }_{\nu} \big\rangle , \quad
				\big| \tilde{ \Psi }_{\nu} \big\rangle = \prod_{\gamma \neq \nu} \left| \psi_{\gamma} \right\rangle .
			\end{equation}
			
			Subsequently, the cluster's state is updated according to the heat bath algorithm.
			For this, an eigenvalue problem is solved:
			
			\begin{equation}
				\hat{H}_{\nu} \left| \psi_{\nu,k} \right\rangle 
				= \varepsilon_{k} \left| \psi_{\nu,k} \right\rangle , \quad k = 1, \ldots 4 ,
			\end{equation}
			
			and based on the obtained spectrum $\left\lbrace \varepsilon_{k} \right\rbrace $, a distribution function is constructed:
			
			\begin{equation}
				F(n) = \sum_{k=1}^{n} p_k , \quad
				p_{k} = \frac{e^{- \beta\varepsilon_{k} }}{\sum_{i} e^{- \beta\varepsilon_{i} }} .
			\end{equation} 
			
			Based on the inequality $F(n-1) < \xi \leqslant F(n)$,
			where $\xi$ is a random variable uniformly distributed over the unit interval,
			a new state $\left| \psi_{\nu,n} \right\rangle$ for cluster $\nu$ is determined.
			Next, utilizing the wave function \eqref{eq:Psi} corresponding to the current system state, we compute the necessary quantum-mechanical averages and collect statistics over MC steps in the standard manner.
			
			The structure of $\hat{H}_{\nu}$ corresponds to the Hamiltonian of the ideal system in MFA; however, the heat bath algorithm explicitly accounts for the inhomogeneous nature of the molecular fields, which are determined by the current state of the system.
			This allows for a more accurate treatment of temperature fluctuations and results in lower critical temperatures of phase transitions compared to MFA predictions.
			Additionally, numerical calculations reveal the instability of homogeneous AFM and P phases and the existence of phase-separation (PS) regions. 
			The ability to determine the phase boundaries of this region is a distinctive feature of the employed MC method.
			
\begin{figure}
				\centering
				\includegraphics[width=\linewidth]{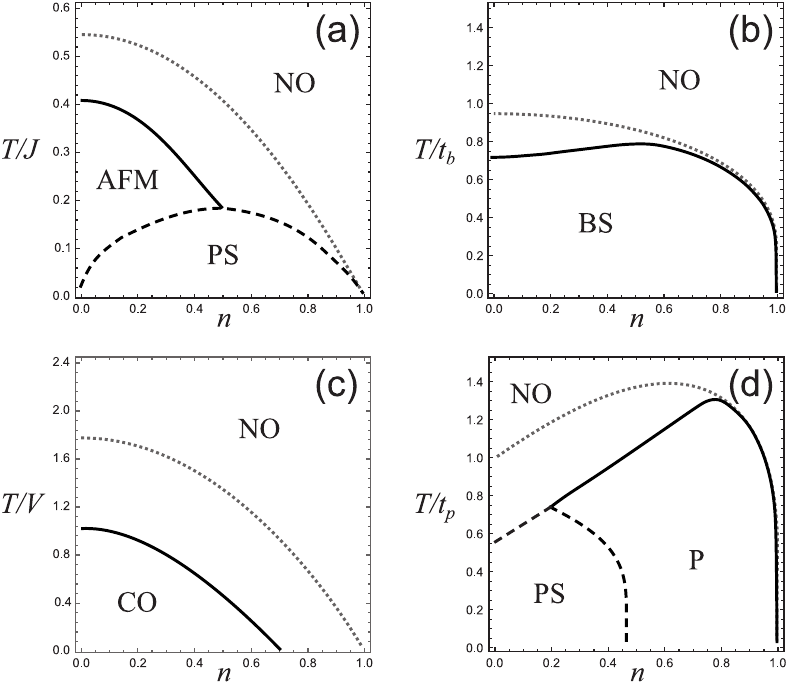}
				\caption{
					Comparison of the results from MFA and numerical modeling using the heat bath algorithm.
					The boundary between the non-ordered (NO) and corresponding ordered phases is shown by a dashed line in MFA, and by a solid line in the heat bath algorithm.
					The boundary of the PS region is indicated by a dotted line.
					The values of the non-zero model parameters are: $\Delta=0.1$, 
					(a) $J=1$;
					(b) $t_{b}=1$;
					(c) $V=1$; 
					(d) $t_{p}=1$.}
				\label{fig:MFA-heatbath}
			\end{figure}

			A comparison of the results from MFA and numerical modeling using the heat bath algorithm is shown in Figure~\ref{fig:MFA-heatbath}.
			A limitation of the algorithm is the high computational cost of an elementary MC step, associated with the necessity of solving an eigenvalue problem at each step.
			
			In the simulation, the parameter $J$ in the Hamiltonian (\ref{eq:Ham0}) was set to 1.
			Consequently, the physical model under consideration possesses six independent parameters.
			To simplify the prediction task, we restricted the analysis to four parameters: $\Delta$, $V$, $t_b$, and $t_p$, given the weak influence of the parameters $t_n$ and $t_{pn}$ on the phase diagram~\cite{Moskvin2020PSS}.

			\section*{Selecting a machine learning model}
			
			Predicting model parameters from spatial distributions is a classic regression problem. To address this, we initially selected two deep learning architectures known for handling complex spatial dependencies: VGG~\cite{simonyan2014very} and ResNet~\cite{he2016deep}.
			
			First, we employed the VGG16BN architecture, a modification of VGG16 incorporating batch normalization (BN) layers to stabilize training and accelerate convergence. 
			Its sequential application of $3 \times 3$ convolutions efficiently captures local spatial features within an image.
			Each convolutional block is followed by a pooling layer, which progressively reduces the spatial dimensions while increasing the number of filters. 
			With 16 weight layers (13 convolutional and 3 fully connected), VGG16 requires a large number of parameters, making it computationally expensive. Furthermore, the aggregation of spatial information in fully connected layers can disrupt fine-grained details, potentially hindering accurate regression. 
			To mitigate these limitations, we also evaluated two variants of the ResNet architecture: ResNet18 and ResNet50. 
			A defining feature of ResNet is the use of skip connections, which alleviate the vanishing gradient problem and enable the effective training of deep networks. 
			The fundamental component, the residual block, adds the input directly to the output of the convolutional layers. 
			ResNet18 is a comparatively lightweight model suitable for problems with limited computational resources, while ResNet50 offers higher performance through bottleneck blocks that optimize depth and parameter efficiency.
			
			Preliminary results indicated that the VGG and ResNet models performed poorly for our specific task. 
			We attribute this to the fact that these models were originally designed for classification, which often does not require the preservation of detailed spatial context. 
			Consequently, we turned to the U-Net architecture, adapting it for the regression task.
			
			U-Net was originally developed for biomedical image segmentation tasks, where pixel-level accuracy is critical~\cite{UNET2015,azad2024medical}.
			Over time, its application has expanded beyond medicine, as evidenced by successful implementations in condensed matter physics~\cite{lee2023device,ye2023u,kamali2024physics,chen2024using}. 
			Previous work also includes examples of U-Net adapted for regression tasks~\cite{yao2018pixel,gertsvolf2024u,lee2023device,chen2024using}.
			
\begin{figure}
				\centering
				\includegraphics[width=\linewidth]{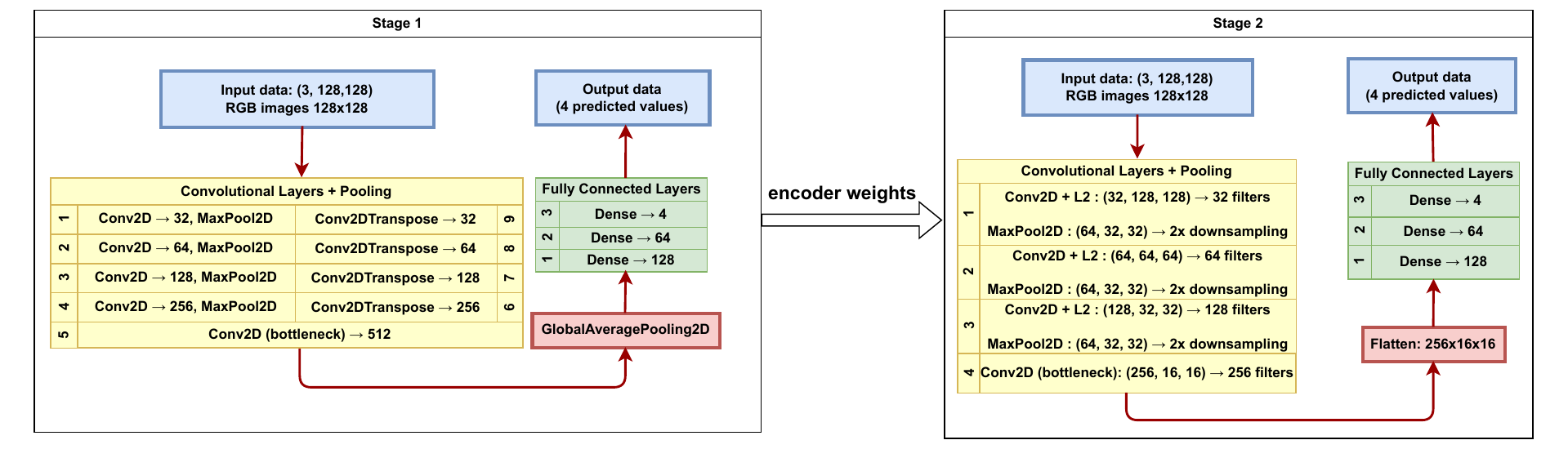}
				\caption{Two-stage neural network operation diagram.}
				\label{fig:unet_scheme}
			\end{figure}

			In this study, we propose a modified U-Net architecture combined with a two-stage training scheme. 
			The operating diagram is shown in Figure~\ref{fig:unet_scheme}. 
			In the first stage, a model with an U-Net-like architecture is pre-trained to operate as an autoencoder, allowing for the identification of common features and patterns in phase diagrams. 
			In the second stage, transfer learning is performed using frozen encoder weights, followed by further training of the regression head with L2 regularization. 
			For this stage, the standard U-Net output layers are replaced with several fully connected layers that transform the feature maps into the four numerical values of the Hamiltonian parameters (\ref{eq:Ham0}).
			The proposed modification also eliminates skip connections from the original architecture to prevent excessive model complexity and focus on the most informative features essential for the regression task.
			
			The dataset was partitioned with 60\% for training, 20\% for validation, and the remaining 20\% for testing. 
			To comprehensively evaluate the predictive accuracy for the four parameter ($\Delta$, $V$, $t_b$, $t_p$), we employed a set of metrics including MAE (Mean Absolute Error), RMSE (Root Mean Squared Error), and $R^2$ (coefficient of determination). 
			Table~\ref{tab:R2_compare} presents the combined results for the U-Net, VGG, and ResNet models, evaluated using the $R^2$ metric on phase diagrams generated with MFA. 
			The dataset size was consistent across all models, consisting of 10,000 phase diagrams.
			
			\begin{table*}[t]
				\centering
				\begin{tabular}{c|c|c|c|c|c||c}
					\hline
					\diagbox[width=4cm]{Model}{Parameter} &$\Delta$ & $V$ & $t_b$ & $t_p$  &Total & Time, s\\
					\hline
					\hline
					U-Net 
					& 0.4972
					& 0.8714
					& 0.6988
					& 0.8818 
					& 0.7373
					& 210
					\\
					\hline
					VGG16BN 
					& 0.5176
					& 0.9132
					& 0.4807
					& 0.7960 
					& 0.6769
					& 3287
					\\
					\hline
					ResNet18 
					& 0.0164 
					& 0.5982
					& 0.5253
					& 0.6937 
					& 0.4584
					& 112
					\\
					\hline
					ResNet50 
					& 0.0012		
					& -0.2436
					& 0.3474
					& 0.4048
					& 0.1274
					& 320
					\\
					\hline
				\end{tabular}
				\caption{Aggregate results of machine learning of different models by the $R^2$ metric and training time for different model parameters}
				\label{tab:R2_compare}
			\end{table*}
			
			The results for $R^2$ convincingly demonstrate that the U-Net-based architecture is optimal for this task.
			Although VGG16BN achieves comparable prediction quality for certain individual parameters (notably $V$ and $t_p$), U-Net outperforms it on the overall metric while requiring approximately 15 times less training time. 
			Furthermore, the U-Net architecture exhibits better generalization ability with limited training data compared to the more parameter-intensive VGG model. 
			This is a critical advantage when working with smaller datasets typical of resource-intensive phase diagram construction methods.
			
			\section*{Results for the U-Net architecture}
			
			This section presents an analysis of the results of predicting four Hamiltonian model parameters ($\Delta$, $V$, $t_b$, and $t_p$) using the U-Net convolutional neural network.
			
			Two independent methods (MFA and the heat bath algorithm) were employed to obtain a diverse sample of phase diagrams in the temperature-density variables.
			Each diagram corresponds to a specific set of model parameters.
			Primary attention was paid to training on the diagrams obtained using MFA, as this approach enables the rapid generation of extensive datasets.
			Calculating a single phase diagram using MFA is at least two orders of magnitude faster than numerical calculation using the heat bath algorithm with similar computing power.
			The larger MFA dataset facilitates a detailed analysis of the influence of individual parameters on phase diagrams.
			Furthermore, the results of validating the approach on data obtained using the heat bath algorithm with a smaller training dataset are also presented.
			
			\subsection*{Prediction results for MFA phase diagrams}
			
\begin{figure}[t]
				\centering
				\includegraphics[width=\linewidth]{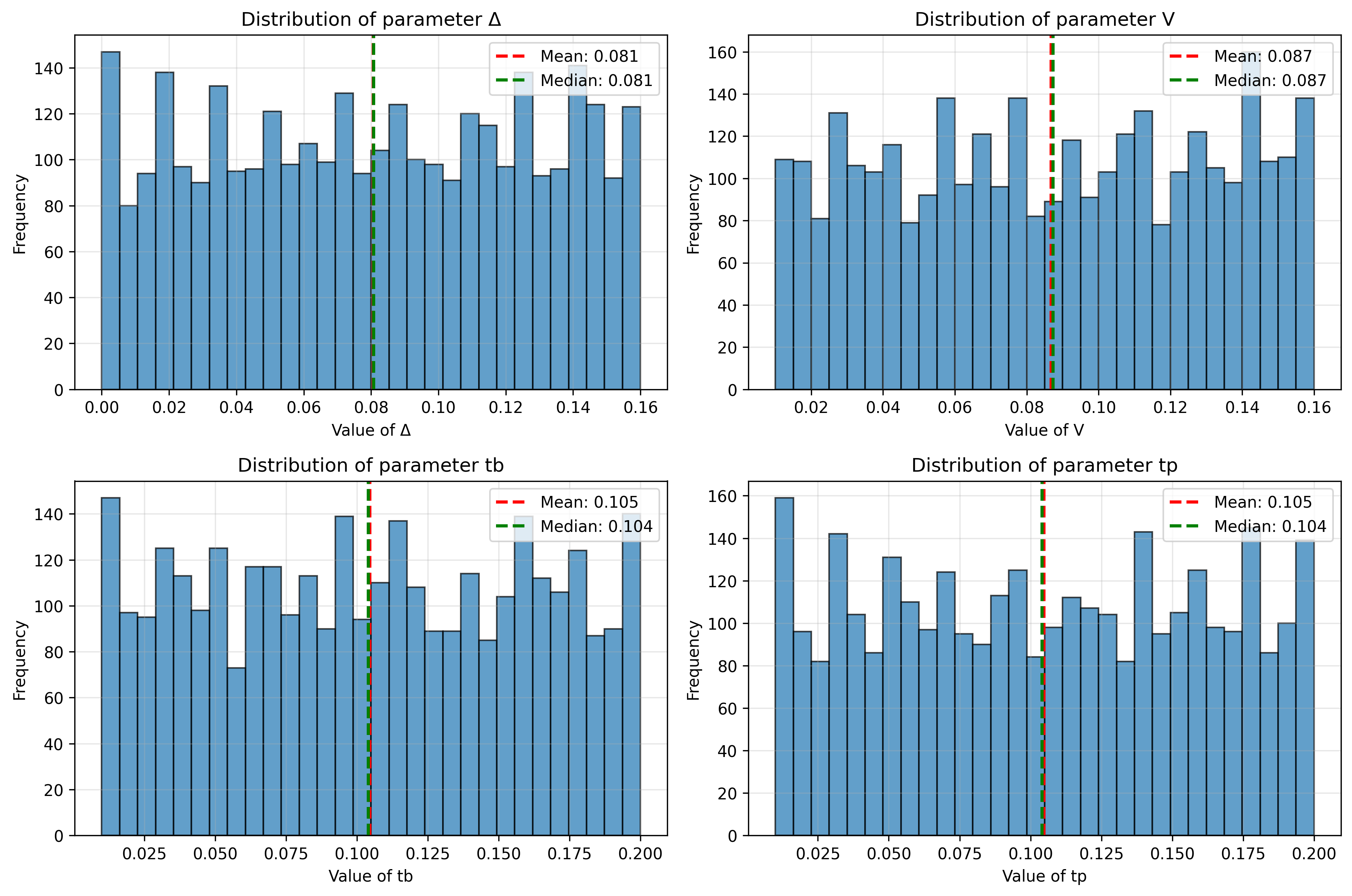}
				\caption{The distribution histograms of the parameters $\Delta$, $V$, $t_b$ and $t_p$ for MFA demonstrate uniform generation of calculation points}
				\label{fig:histo_mfa}
			\end{figure}

			To ensure that the calculation points in the neural network training dataset are uniformly distributed and to avoid data imbalance, we randomly generated new parameter sets based on the previously calculated points in the parameter space.
			As illustrated in Figure~\ref{fig:histo_mfa}, the distributions of all four parameters are characterized by close mean and median, indicating the absence of systematic bias and ensuring data balance. 
			Comparative analysis shows that random uniform parameter distributions yield a 20–30\% increase in prediction accuracy compared to equidistant sampling methods.
			
\begin{figure}
				\centering	
				\begin{minipage}[t]{0.32\linewidth}
					\includegraphics[width=\linewidth]{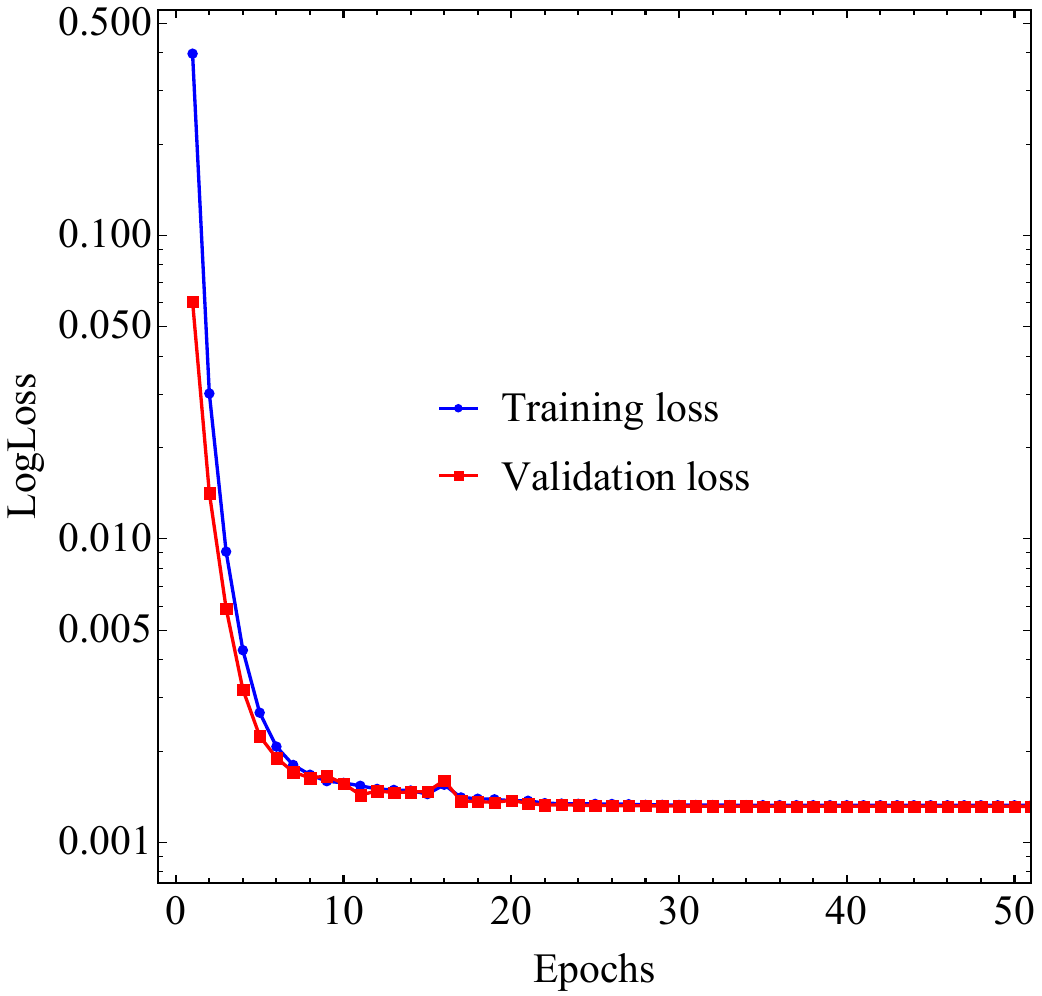}
					\caption{U-Net loss curve for MFA phase diagrams  on a logarithmic scale}
					\label{fig:loss_mfa}
				\end{minipage}
				\hfill
				\begin{minipage}[t]{0.64\linewidth}
					\includegraphics[width=\linewidth]{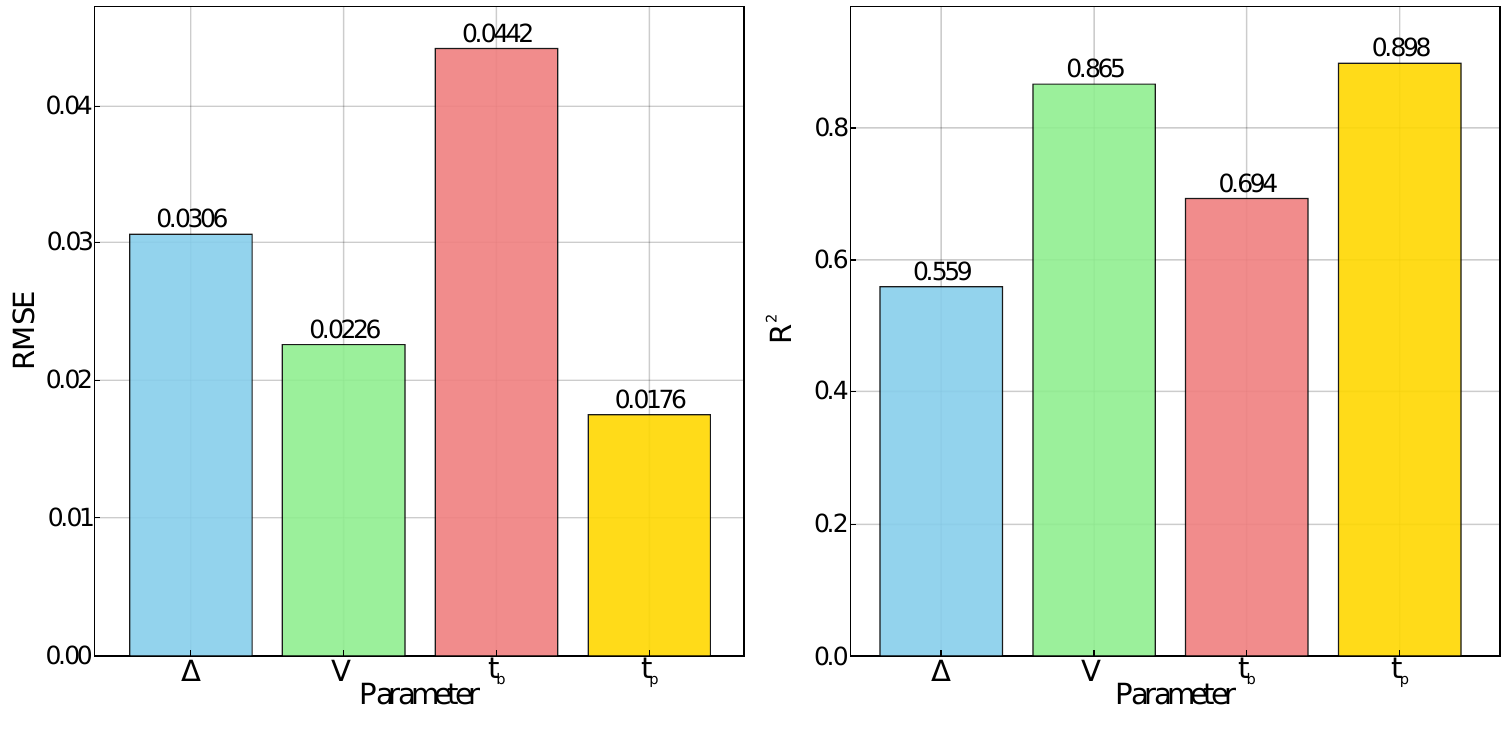}
					\caption{Bar charts of RMSE and $R^2$ metrics for each parameter $\Delta$, $V$, $t_b$, $t_p$ for MFA}
					\label{fig:metrics_mfa}
				\end{minipage}
			\end{figure}

			Figure~\ref{fig:loss_mfa} shows the loss function curves during training and validation on a logarithmic scale.
			These curves converge rapidly and stabilize at very low levels, indicating indicates a stable optimization process and the absence of overfitting.
			
			A qualitative assessment of the prediction accuracy for each parameter is presented in Figure~\ref{fig:metrics_mfa}, which shows bar charts of the key metrics RMSE and $R^2$ for each predicted parameter. 
			Notably, there is significant variability in prediction accuracy between different parameters, indicating various degrees of parameters influence on phase diagrams.
			
\begin{figure}
				\centering
				\includegraphics[width=\linewidth]{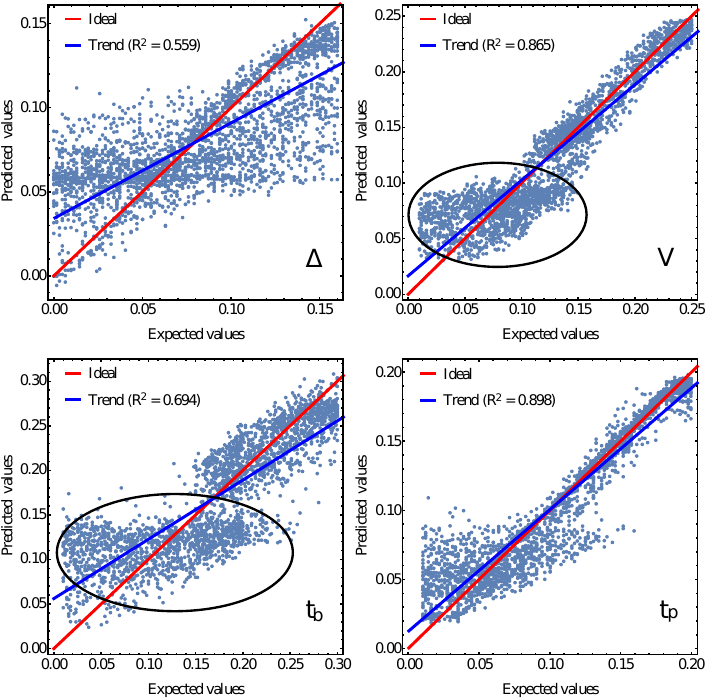}
				\caption{Scatter plots show correlation between predicted and expected values for each parameter for MFA phase diagrams.
					Areas where there is no correlation and the neural network can not predict the parameters are outlined in black}
				\label{fig:scatter_mfa}
			\end{figure}

			For visual evaluation of the correlation between the predicted and expected parameter values, scatter plots are shown in Figure~\ref{fig:scatter_mfa}. The red line indicates the linear dependence, representing the line of best fit. 
			The blue line denotes the main trend according to the coefficient of determination $R^2$. 
			Areas showing no correlation between the predicted and expected parameter values are outlined in black.
			
			For parameters $\Delta$ and $t_p$, the points are mostly cluster along the red ideal predictive line across the entire range of parameter variations, which confirm the high quality of the regression. 
			Scatter plots for $t_b$ and $V$ reveal heterogeneity with near-zero correlation outside the general trend in regions of small parameter values. These regions are highlighted with black ovals in Figure~\ref{fig:scatter_mfa}.
			
			This effect does not necessarily indicate mistakes in the machine learning model. 
			Several factors could account for this.
			First, there might be an imbalance in the training dataset. 
			Although the scatter plots display horizontal regions at certain values of $t_b$ and $V$, the distribution of data points for these parameters was shown to be approximately normal (see Figure~\ref{fig:histo_mfa}), which rules out this explanation.
			
			A second potential reason is an inherent limitation of the physical model: if the parameters $t_b$ or $V$ are weakly correlated with the form of phase diagrams, the model may be unable to learn these relationships effectively. Furthermore, degeneracy of the phase diagrams is possible in the highlighted regions, since different parameter combinations can lead to statistically indistinguishable configurations.
			This can impose fundamental limits on predictive accuracy. 
			To test this hypothesis, we fixed the mean values of three parameters and sequentially varied the fourth one under investigation.
			
\begin{figure*}[t]
				\centering
				\includegraphics[width=\linewidth]{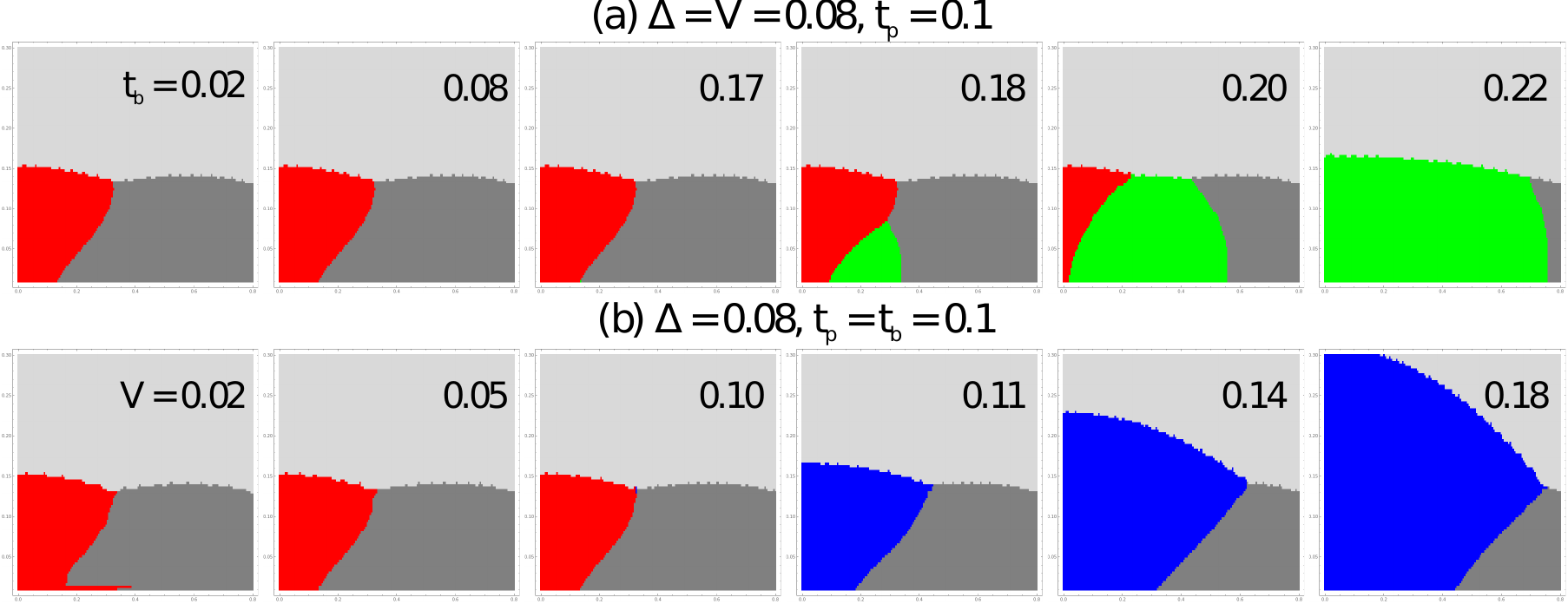}
				\caption{
					Sensitivity analysis of the $T$--$n$ phase diagrams to parameter variations:
					(a) a series of diagrams plotted for fixed $\Delta$, $V$, and varying $t_b \in [0.02 - 0.22]$;
					(b) a series of diagrams plotted for fixed $\Delta$, $t_b$, and varying $V \in [0.02 - 0.18]$.
					Different colors represent different homogeneous phases. 
					Light grey indicates the NO phase, red represents AFM, green represents CO, blue represents BS, and dark grey represents the Fermi liquid phase -- an analogue of the P-phase with simultaneously non-zero order parameters $\big\langle \hat{P}_{m}^{+} \big\rangle$ and $\big\langle \hat{N}_{m}^{+} \big\rangle$
				}
				\label{fig:PD_mfa}
			\end{figure*}

			Series of phase diagrams were calculated with sequential variation of the target parameters:  
			to analyze the influence of parameter $t_b$ in the interval from 0.02 to 0.22 with fixed values of parameters $\Delta$, $V$, and $t_p$ (Figure~\ref{fig:PD_mfa}(a));
			to analyze the influence of parameter $V$ in the interval from 0.02 to 0.18 with fixed values of parameters $\Delta$, $t_b$, and $t_p$ (Figure~\ref{fig:PD_mfa}(b)).
			
			The results of the computational experiment confirmed that the phase diagrams remains nearly identical throughout the entire ranges of zero correlation between the predicted and expected values of parameters.
			This means that variations of $t_b$ and $V$ within these limits does not lead to qualitative or substantial quantitative changes in the phase diagram. 
			This observation elucidates the fundamental limitation of the prediction accuracy: the neural network can not reconstruct parameters that do not significantly affect the resulting phase diagram image.
			However, as demonstrated in Figure~\ref{fig:PD_mfa}, beyond the critical values of parameters ($t_b \ge 0.18$, $V \ge 0.11$), qualitative changes in phase diagrams begin to manifest, which correlates with the emergence and strengthening of the statistical relationship in the scatter plots in Figure~\ref{fig:scatter_mfa}.
			
			This suggests that the machine learning model acts not merely as a ``black box'', but as a tool capable of analyzing the physical properties of the underlying physical model.
			The neural network's inability to predict certain parameters indicates their weak influence on the target variable (the phase diagram) within the considered range.
			Conversely, it successfully identifies parameter intervals where the parameter becomes important and significantly affects the form of the diagram.
			
			Low prediction accuracy in specific ranges indicates parametric degeneracy within the system, whereas high accuracy (error margins of 1–3\%) is achieved in regions where the parameter significantly influences the morphological characteristics of the phase diagram.
			
			\subsection*{Prediction results for the heat bath algorithm phase diagrams}
			
\begin{figure}
				\centering
				\includegraphics[width=\linewidth]{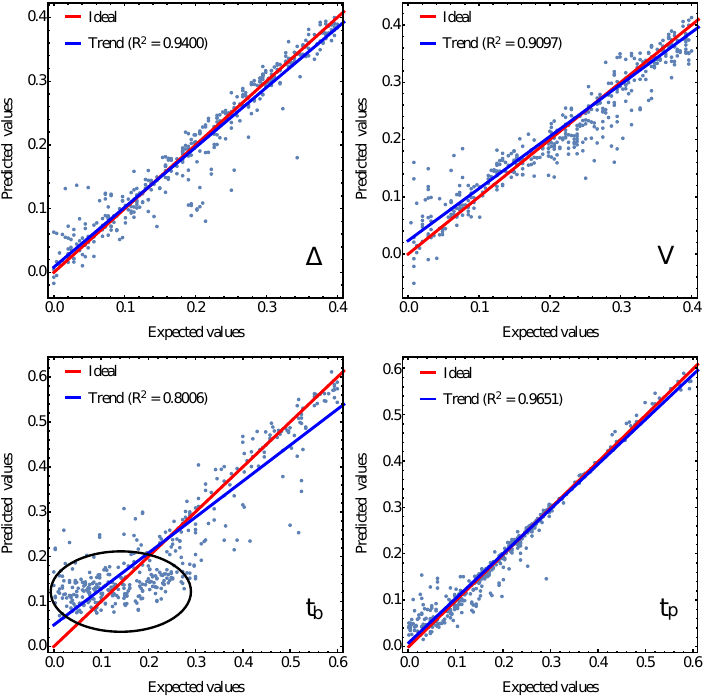}
				\caption{
					Scatter plots of the predicted and expected parameter values for the heat bath algorithm. 		
					For small $t_b$, the lack of predictive power of the neural network is demonstrated due to the independence of phase diagrams on $t_b$. However, in other regions, the determination coefficient $R^2$ demonstrates excellent convergence}
				\label{fig:scatter_cluster1}
			\end{figure}

			Using a more precise, but computationally intensive heat bath algorithm, we calculated a dataset of 1,200 phase diagrams.
			Figure~\ref{fig:scatter_cluster1} presents the correlation scatter plots, which demonstrate excellent coefficients of determination $R^2$ for each parameter, as well as the presence of a significant region of low $t_b$ values with zero correlation between the predicted and expected values.
			
			Subsequent parametric analysis again revealed that, within the considered range of values, the phase diagrams also exhibited weak sensitivity to this parameter.
			This indicates that, within a more rigorous approach, the effect of the $t_b$ parameter on the observed phase diagrams within the same range becomes less pronounced or is masked by other effects.
			Consequently, the primary source of prediction errors is again not a deficiency of the machine learning model but rather the physical insignificance of the parameter within the specified range.
			This underscores the importance of accurately defining the boundaries of parameter space during planning stages of computational experiments.
			
			\subsection*{Comparison of results for two approaches}
			
			Table 1 shows the prediction quality metrics (RMSE and $R^2$) for the best neural network predictions based on the U-Net architecture for each of the four model parameters.
			These results pertain to the largest training datasets and are evaluated for both the MFA approach and the heat bath algorithm.
			
			It is important to note that the calculation was performed exclusively within regions of the parameter space with pronounced correlation (excluding the regions of insensitivity outlighted in black in Figures~\ref{fig:scatter_mfa} and~\ref{fig:scatter_cluster1}).
			
			\begin{table*}[t]
				\centering
				\begin{tabular}{c||c|c|c|c}
					\diagbox[width=4cm]{Model}{Parameter} & $\Delta$ & $V$ & $t_b$ & $t_p$ \\
					\hline
					\hline
					MFA 
					& 0.016 $\vert$ 0.829
					& 0.013 $\vert$ 0.937
					& 0.031 $\vert$ 0.967
					& 0.014 $\vert$ 0.939 \\
					\hline
					Heat bath 
					& 0.028 $\vert$ 0.940 
					& 0.034 $\vert$ 0.919 
					& 0.033 $\vert$ 0.861 
					& 0.025 $\vert$ 0.965 \\
					\hline
				\end{tabular}
				\caption{Best U-Net prediction quality metrics (RMSE $\vert$ $R^2$) for different model parameters and various methods to obtain phase diagrams}
			\end{table*}
			
			Thus, the machine learning model serves as a universal tool for identifying physically significant parametric regions and demonstrates consistent results for different computational methods.
			The observed limitations in predictive capability reflect the fundamental properties of the physical system rather than architectural deficiencies of the U-Net model.
			This confirms the validity of this approach for analyzing the parametric sensitivity of complex multi-parametric systems, including those with quantum effects.
			
			A promising research direction involves investigating transfer learning between different computational methods. 
			Pre-training the model on extensive MFA datasets followed by fine-tuning on limited heat bath algorithm data could significantly enhance learning efficiency and prediction accuracy. 
			This approach would offset the computational cost of more accurate but complex methods by using fast approximate calculations for model initialization. 
			Cross-method transfer learning research is especially relevant for problems where obtaining more accurate data is computationally constrained.
			
			\section*{Conclusions}
			
			This study demonstrates the effectiveness of machine learning for solving the inverse problem of reconstructing the parameters of a model Hamiltonian for a cuprate superconductor Hamiltonian based on its phase diagram.
			
			We considered a pseudospin model of cuprates, where a pseudospin $S=1$ describes the three multi-electron states of the [CuO$_4$]$^{7-,6-,5-}$ centers.
			The complex multi-parametric Hamiltonian includes local and non-local density-density correlations for [CuO$_4$]$^{7-,5-}$ centers, Heisenberg exchange interaction for magnetic [CuO$_4$]$^{6-}$ centers, as well as three types of correlated single-particle transfers and a two-particle transfer.
			Phase diagrams in the temperature--density variables were obtained using two approaches: mean-field approximation and heat bath Monte Carlo algorithm.
			
			Each parameter of the model Hamiltonian influences the resulting phase diagrams in a complex and highly sensitive manner. 
			This complexity makes the inverse problem of precisely tuning parameters to reproduce experimental results an extremely resource-intensive task.
			
			To address this inverse problem, three deep learning architectures were considered: VGG, ResNet, and U-Net.
			A comparative analysis revealed a clear advantage of U-Net architecture both in prediction accuracy and computational efficiency.
			This is attributed to the ability of U-Net to preserve spatial context through its skip-connection mechanism, which is proved crucial for the analysis of phase diagrams.
			
			The use of a U-Net architecture adapted for regression tasks enabled high-accuracy predictions across all considered model parameters.
			Analysis of regions with low correlations between predicted and expected parameters revealed fundamental limitations in predictive capabilities, which are not related to shortcomings of the machine learning model, but rather to physical properties inherent to the system.
			
			Areas of low prediction accuracy for parameters correspond to regions of parametric insensitivity in the phase diagrams.
			This indicates that neural networks can serve as effective tools for identifying the significance of model parameters on the system.
			This was confirmed by parametric analysis for both the mean-field approximation and the heat bath Monte Carlo algorithm.
			The latter demonstrated a greater insensitivity to the two-particle transport parameter across a wide parameter range.
			Thus, the U-Net neural network proved itself not merely as a ``black box'' but as a tool for verifying the physical properties of the system.
			
			The model demonstrated high performance across different computational methods.
			Despite a significantly smaller training dataset for the heat bath algorithm data, the U-Net maintained high prediction accuracy ($R^2$ > 0.9 for most parameters), which confirms the reliability of the proposed approach and its applicability to real computational data.
			
			Future prospects for this work involve extending the model to incorporate additional parameters, utilizing more complex neural network architectures, investigating transfer learning between different computational methods, and extending the approach to other multi-parametric physical models, including complex quantum systems with strong correlations.
			The proposed method opens up the possibility of automated selection of theoretical model parameters based on experimental phase diagrams.
			This is particularly significant in the light of increasing computational complexity of modern physical models.
			
			\section*{Acknowledgments}
			The research was supported by the Russian Science Foundation, grant no. 24-21-20147.
			
			\bibliographystyle{elsarticle-num}
			\bibliography{references}
			
	\end{document}